\documentclass[review,number,sort&compress]{elsarticle}
\usepackage{lineno}

\usepackage{amssymb}

\journal{Nucl.\ Instrum.\ Methods Phys.\ Res.\ A}

\begin{document}

\begin{frontmatter}



\title{Comparison of magnetic field uniformities for discretized and
finite-sized standard $\cos\theta$, solenoidal, and spherical coils}


\author{N.\ Nouri}
\author{B.\ Plaster}
\address{Department of Physics and Astronomy, University of Kentucky, \\
Lexington, Kentucky 40506, USA}

\begin{abstract}
A significant challenge for experiments requiring a highly uniform
magnetic field concerns the identification and design of a discretized
and finite-sized magnetic field coil of minimal size.  In this work we
compare calculations of the magnetic field uniformities and field
gradients for three different standard (i.e.,
non-optimized) types of coils: $\cos\theta$,
solenoidal, and spherical coils.  For an experiment with a particular
requirement on either the field uniformity or the field gradient, we
show that the volume required by a spherical coil form which satisfies
these requirements can be significantly less than the volumes required
by $\cos\theta$ and solenoidal coil forms.
\end{abstract}



\end{frontmatter}

\section{Introduction}
\label{sec:intro}
\begin{figure}
\begin{center}
\includegraphics[scale=0.45]{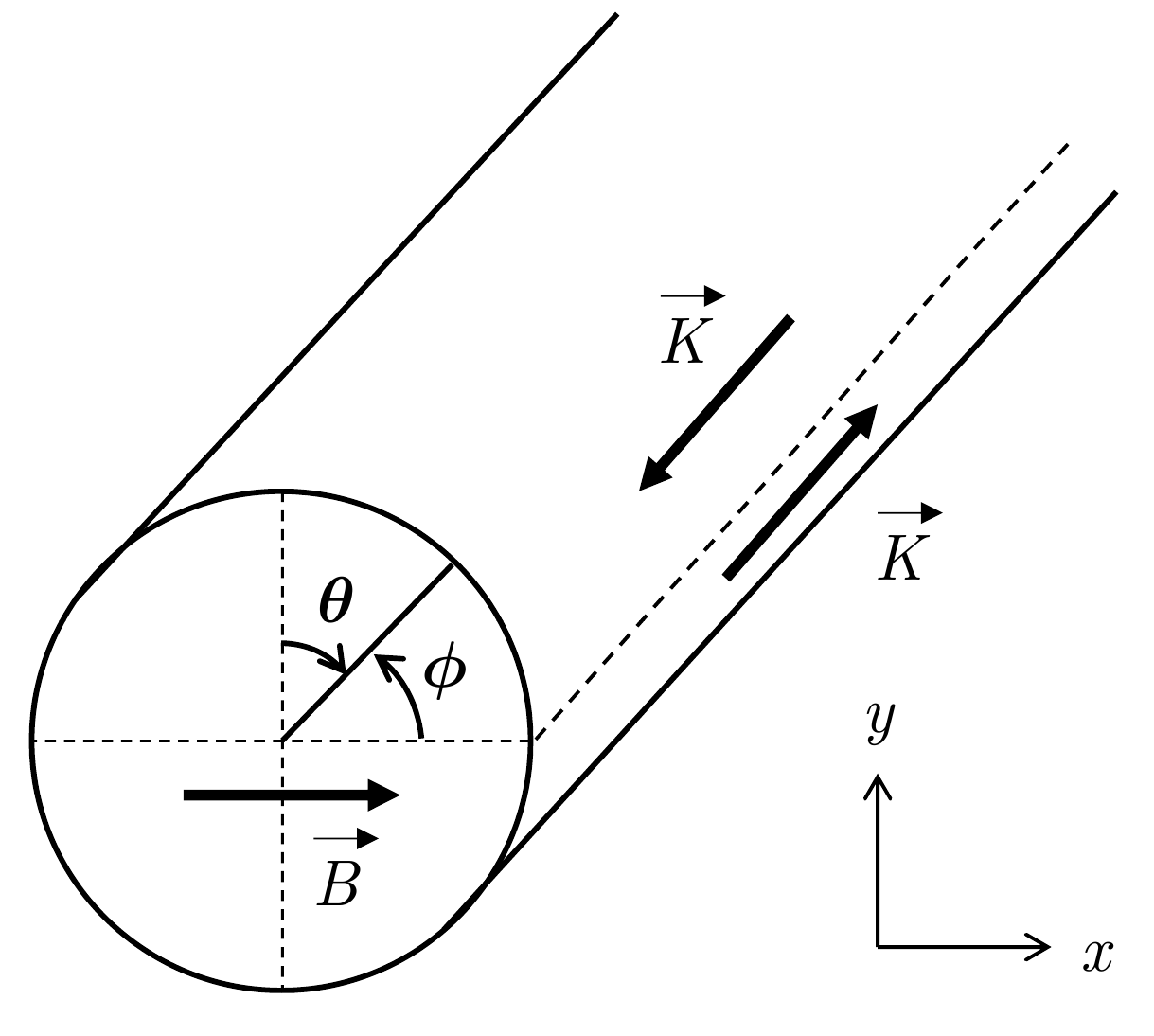}
\caption{Schematic diagram of a $\cos\theta$ coil with a continuous
  surface current distribution $\vec{K} = K \cos\theta\hat{z}$.
  Note that a coil with such a surface current distribution is also
  commonly referred to as a ``sine-phi'' coil \cite{jin98,bidinosti05},
  where the angle $\phi$ is the usual cylindrical coordinate.}
\label{fig:cos_theta_coil_schematic}
\end{center}
\end{figure}

As is well known \cite{jacksonEM,smythe}, a solenoid of infinite
length will generate a perfectly uniform axial magnetic field
everywhere inside of a cylindrical volume.  As is also well known
\cite{jacksonEM,smythe}, an infinitely-long ``$\cos\theta$ coil''
(i.e., a cylindrical coil with a continuous surface current $\vec{K}
= K \cos\theta \hat{z}$, with the angle $\theta$ defined relative
to the $\hat{y}$-axis as shown in
Fig.\ \ref{fig:cos_theta_coil_schematic}), will generate a perfectly
uniform transverse magnetic field (along the $\hat{x}$-direction in
Fig.\ \ref{fig:cos_theta_coil_schematic}) within a cylindrical volume.
And indeed, experiments requiring a highly uniform magnetic field have
generally employed discretized and finite-length solenoidal or
$\cos\theta$ coils for the generation of highly uniform axial or
transverse magnetic fields.  For example, experimental searches for
non-zero permanent electric dipole moments, such as of the neutron and
the $^{199}$Hg and $^{225}$Ra atoms, require highly uniform magnetic
fields and past and future experiments
\cite{baker06,balashov07,altarev09,griffith09,holt10,perezgalvan11}
have typically used highly-optimized $\cos\theta$ or solenoidal coils.
In these experiments, the volume required by the $\cos\theta$ and
solenoidal cylindrical coil forms is, in general, significantly larger
than the sensitive experimental volume over which the requirements on
the field uniformity and/or field gradient must be satisfied.

\begin{figure}
\begin{center}
\includegraphics[scale=0.45]{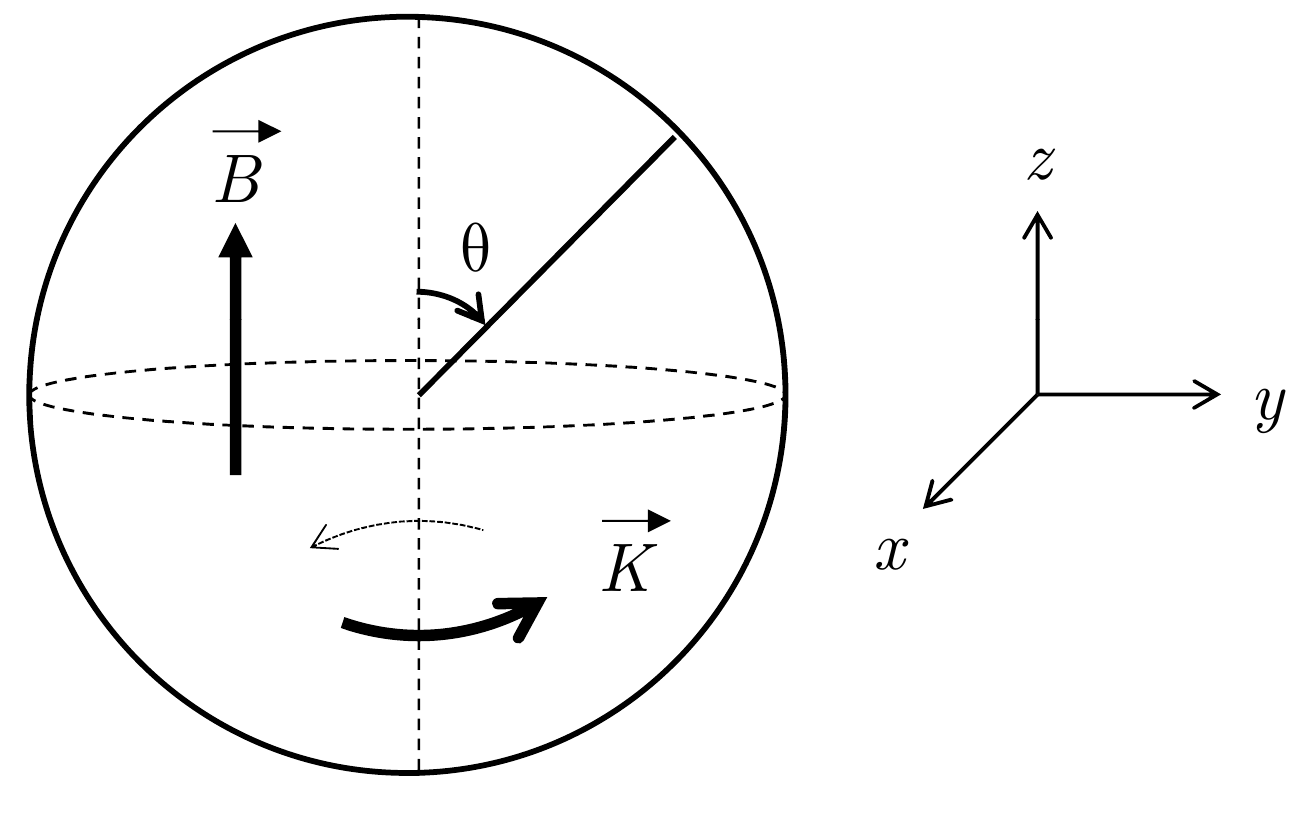}
\caption{Schematic diagram of a spherical coil with a continuous
surface current distribution $\vec{K}=K \sin\theta\hat{\phi}$.}
\label{fig:spherical_coil_schematic}
\end{center}
\end{figure}

Perhaps less well known or appreciated is the fact that a
\textit{finite-sized} spherical coil with a continuous surface current
$\vec{K} = K\sin\theta \hat{\phi}$, shown schematically in
Fig.\ \ref{fig:spherical_coil_schematic}, where $(\theta,\phi)$ are
the usual spherical coordinates, will generate a perfectly uniform
magnetic field along the sphere's $\hat{z}$-axis within the spherical
volume \cite{jacksonEM,smythe,haus_melcher,everett66,jin98}.  Although
the technical challenges associated with the fabrication of an
experimentally-realizable discretized spherical coil (e.g., precise
wire placement, etc.) may be more significant than those associated
with the fabrication of solenoidal and discretized $\cos\theta$ coils,
we note that the potential appeal of a spherical coil is that even for
the (non-realistic) case of continuous surface currents, a spherical
coil of finite size will generate a perfectly uniform field, whereas
the $\cos\theta$ and solenoidal coils must be of infinite length.
That is, a spherical coil does not suffer from any such
``finite-length'' effects'' that are by definition present in
finite-sized $\cos\theta$ and solenoidal coils.  Therefore, we
conjecture that the ratio of the fiducial volume (i.e., the volume
over which the requirements on the field uniformity or field gradient
must be satisfied) to the total volume occupied by a spherical coil
form will be larger than this ratio for $\cos\theta$ and solenoidal
coil forms.  And, indeed, a new experimental search for the neutron
electric dipole moment will employ a spherical coil \cite{masuda12}.
However, to our knowledge, detailed calculations of the off-axis field
uniformity properties of a discretized spherical coil have not been
presented in the literature.

In the remainder of this paper we investigate this conjecture.  We
begin, in Section \ref{sec:models}, by describing our models for our
calculations of the magnetic fields for these three coil types,
including our methods for the calculation of the off-axis fields for
the solenoidal and spherical coils.  We then present the numerical
results of our calculations in Section \ref{sec:results}, where we
ultimately compare the ratios of the fiducial volume to the total
volume occupied by each of the coil types for various requirements on
the field uniformity or field gradient.  Finally, we conclude with a
summary of our findings in Section \ref{sec:summary}.  We emphasize
that throughout this paper we are considering standard (i.e.,
non-optimized) versions of the $\cos\theta$ and solenoidal coils
(i.e., truncated versions of the infinitely-long ideal coils).

\section{Models for Discretized and Finite-Sized Coils}
\label{sec:models}

In this section we describe our numerical models for the calculation
of the magnetic fields (on- and off-axis) from discretized and
finite-sized $\cos\theta$, solenoidal, and spherical coils.  Note that
our discretization of the ideal surface currents is such that we use
a single point (i.e., zero radius) line wire to approximate regions
of equal integrated surface current.  We describe our implementation
of such for each of the coil types below.

\subsection{Model for $\cos\theta$ Coil}
\label{sec:models_cos_theta}

\begin{figure}
\begin{center}
\includegraphics[scale=0.50]{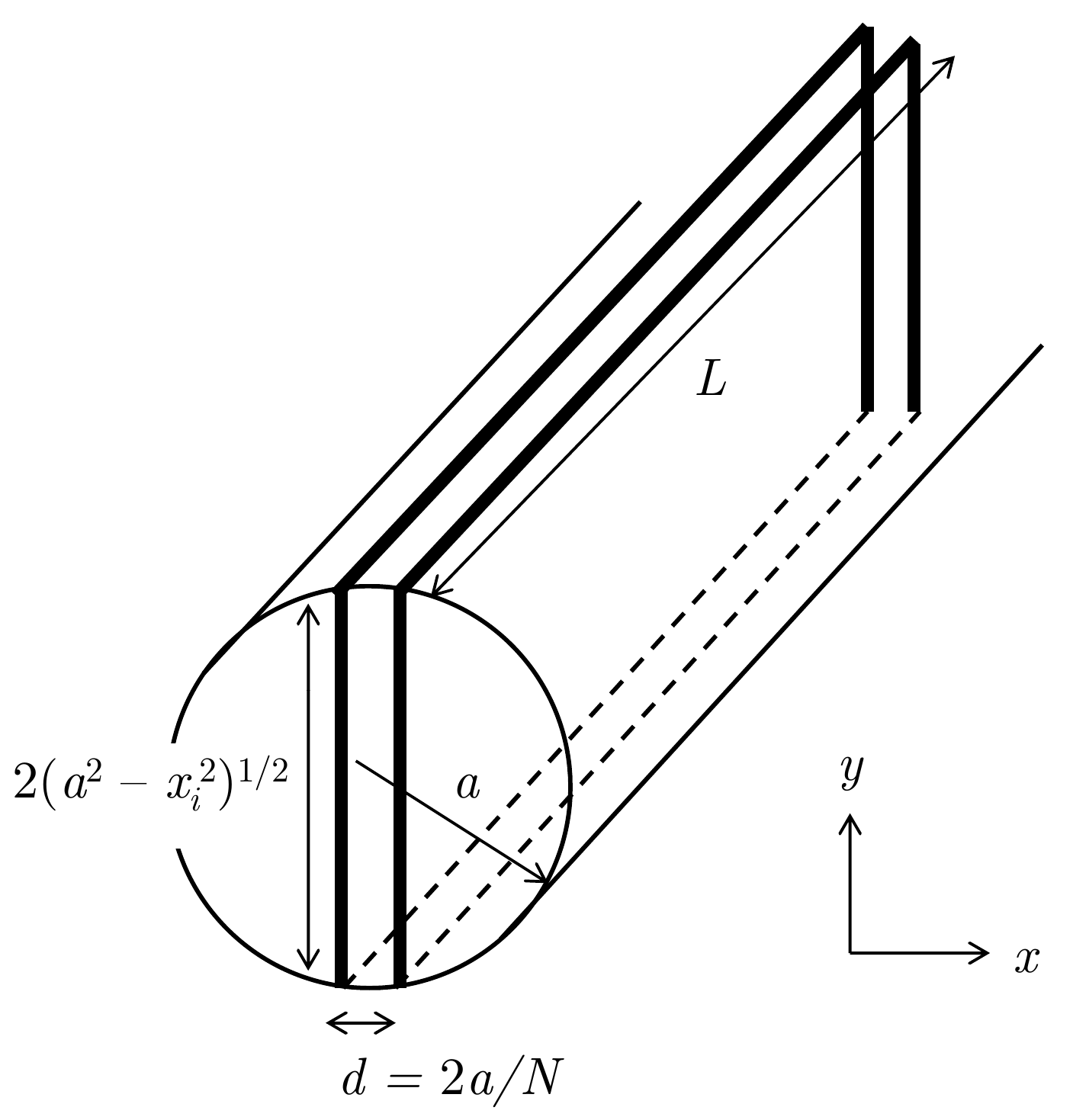}
\caption{Schematic diagram of our model for a standard discretized and
  finite-sized $\cos\theta$ coil.  The thick lines indicate the
  rectangular current loops.  Note that this choice of a return path
  for the currents is not unique; another well-known approach is to
  employ a ``saddle-shaped'' winding \cite{bidinosti05,perezgalvan11},
  in which the return wires are wound around the perimeters of the
  cylinder ends.}
\label{fig:cos_theta_coil_discretized}
\end{center}
\end{figure}

A schematic diagram of our discretized and finite-length $\cos\theta$
coil is shown in Fig.\ \ref{fig:cos_theta_coil_discretized}.  As shown
in Ref.\ \cite{bidinosti05}, the discretization of an infinitely-long
continuous $\vec{K}\ = K \cos\theta\hat{z}$ surface current
distribution into line currents approximating the continuous surface
current distribution (obtained by apportioning the total integrated
current from $\theta = 0$ to $\pi/2$ into equal discrete parts) is
such that the line currents are spaced at equal intervals along the
indicated $x$-axis.  Thus, as indicated in
Fig.\ \ref{fig:cos_theta_coil_discretized}, our model consists of a
cylindrical shell of radius $a$ and length $L$, upon which are wound
$N$ rectangular current loops, spaced at equal intervals $d = 2a/N$
along the $x$-axis at positions $x = \pm \frac{1}{2}d$, $\pm
\frac{3}{2}d$, \ldots, $\pm \frac{N-1}{2}d$.  The line currents then
flow in the $+z$ ($-z$) direction for $\cos\theta>0$ $(<0)$ along the
length of the cylinder.  Our model then assumes a simple return path
for the currents (i.e., straight line currents along the $\pm y$
direction), such that our closed current loops are rectangular.
Thus, all of the
rectangular current loops are of the same length $L$ along the axis of
the cylindrical form (i.e., the $z$-axis), but are of different
lengths along the $y$-axis, where these lengths are constrained by the
cylindrical form to be $2\sqrt{a^2 - x_i^2}$, where $x_i$ denotes the
$i$th loop's position along the $x$-axis.

The calculations we present later have assumed $N$ even; one could, of
course, consider $N$ odd, with an additional wire located at
$\theta=0$, but this would not modify the symmetry of the problem.
Also, note that although our model does not include any connecting
wires between adjacent current loops (which are, of course, required
if a $\cos\theta$ is to be wound with a single wire), an actual
$\cos\theta$ coil can be wound in such a way so that the contributions
to the field from the connecting wires carrying the current from, say,
loop $i$ to loop $i+1$ to loop $i+2$, etc.\ are largely cancelled by
the return wire to the current source.

It is then straightforward to calculate the magnetic field everywhere
in space in rectangular coordinates via the Biot-Savart law.

\subsection{Model for Solenoidal Coil}
\label{sec:models_solenoidal}

Because the surface current distribution on an ideal solenoid is
uniform in the cylindrical coordinates $\phi$ and $z$ (with $z$
oriented the solenoid axis), such that $\vec{K} = K \hat{\phi}$,
our model for a discretized and
finite-length solenoidal coil consists of a cylindrical coil form of
radius $a$ and length $L$ upon which are wound $N$ circular current
loops of radius $a$ oriented in the indicated $xy$-plane and spaced at
equal intervals $d$ along the length of the cylinder.  Note that each
of our discretized current loops is ``isolated'' (i.e., stand-alone
circular loops, with no connecting wires), such that we do not attempt
to account for the effects of a helical winding inherent to the
winding of a solenoid with one continuous wire.

We then employ cylindrical coordinates to calculate the magnetic field
of our solenoidal coil at some observation point $\vec{x} =
(\rho,\phi,z)$.  By symmetry, the vector potential
$\vec{A}_i(\vec{x})$ of the $i$th circular current loop centered at
$(0,0,z_i)$ includes only a $\phi$-component, and is of the form
\cite{jacksonEM,smythe}
\begin{equation}
A_{\phi,i}(\rho,z) = \frac{\mu_0 I}{\pi \kappa_i} \sqrt{\frac{a}{\rho}}
\left[\left(1 - \frac{1}{2}\kappa_i^2\right)K(\kappa_i) - E(\kappa_i)\right],
\end{equation}
where $K(\kappa_i)$ and $E(\kappa_i)$ denote, respectively, the
complete elliptic integrals of the first and second kind, $\kappa_i^2
\equiv 4a\rho/[(a+\rho)^2 + (\Delta z_i)^2]$, and $\Delta z_i \equiv z
- z_i$.

The resulting $\rho$- and $z$-components of the $\vec{B}_i(\vec{x})$
field at $\vec{x}$ due to the $i$th circular current loop are then
calculated as $\vec{B} = \vec{\nabla} \times \vec{A}$ in cylindrical
coordinates and are of the form
\begin{eqnarray}
B_{\rho,i}(\vec{x}) &=& \frac{\mu_0 I (\Delta z_i)}
{2\pi\rho\sqrt{(a+\rho)^2 + (\Delta z_i)^2}}
\Bigg[ -K(\kappa_i)
+ \frac{a^2 + \rho^2 + (\Delta z_i)^2}
{(a-\rho)^2 + (\Delta z_i)^2}E(\kappa_i)\Bigg], \nonumber \\
\\
B_{z,i}(\vec{x}) &=& \frac{\mu_0 I}{2\pi\sqrt{(a+\rho)^2 + (\Delta z_i)^2}}
\Bigg[K(\kappa_i)
+ \frac{a^2 - \rho^2 - (\Delta z_i)^2}
{(a-\rho)^2 + (\Delta z_i)^2}E(\kappa_i)\Bigg], \nonumber
\end{eqnarray}
with the complete elliptic integrals then expressed as power series
\cite{arfken}
in the parameter $\kappa_i$,
\begin{eqnarray}
K(\kappa_i) &=& \frac{\pi}{2}\left\{1 + \sum_{n=1}^\infty
\left[\frac{(2n-1)!!}{(2n)!!}\right]^2 \kappa_i^{2n}\right\}, \nonumber \\
\\
E(\kappa_i) &=& \frac{\pi}{2}\left\{1 - \sum_{n=1}^\infty
\left[\frac{(2n-1)!!}{(2n)!!}\right]^2 \frac{\kappa_i^{2n}}{2n-1}\right\}.
\nonumber
\end{eqnarray}
Note that the values of $\kappa_i$ are restricted to the range $0 \leq
\kappa_i \leq 1$.  We sum the power series until the contribution of
the $n$th term is $< 10^{-9}$.  The resulting $B_{\rho,i}(\vec{x})$
and $B_{z,i}(\vec{x})$ field components for each circular current loop
are then converted to rectangular field components.  The total field
$\vec{B}(\vec{x})$ at observation point $\vec{x}$ from all $N$
circular current loops comprising the discretized solenoidal coil is
then simply the superposition of the rectangular field components from
each of the current loops.

\subsection{Model for Spherical Coil}
\label{sec:models_spherical}

\begin{figure}
\begin{center}
\includegraphics[scale=0.50]{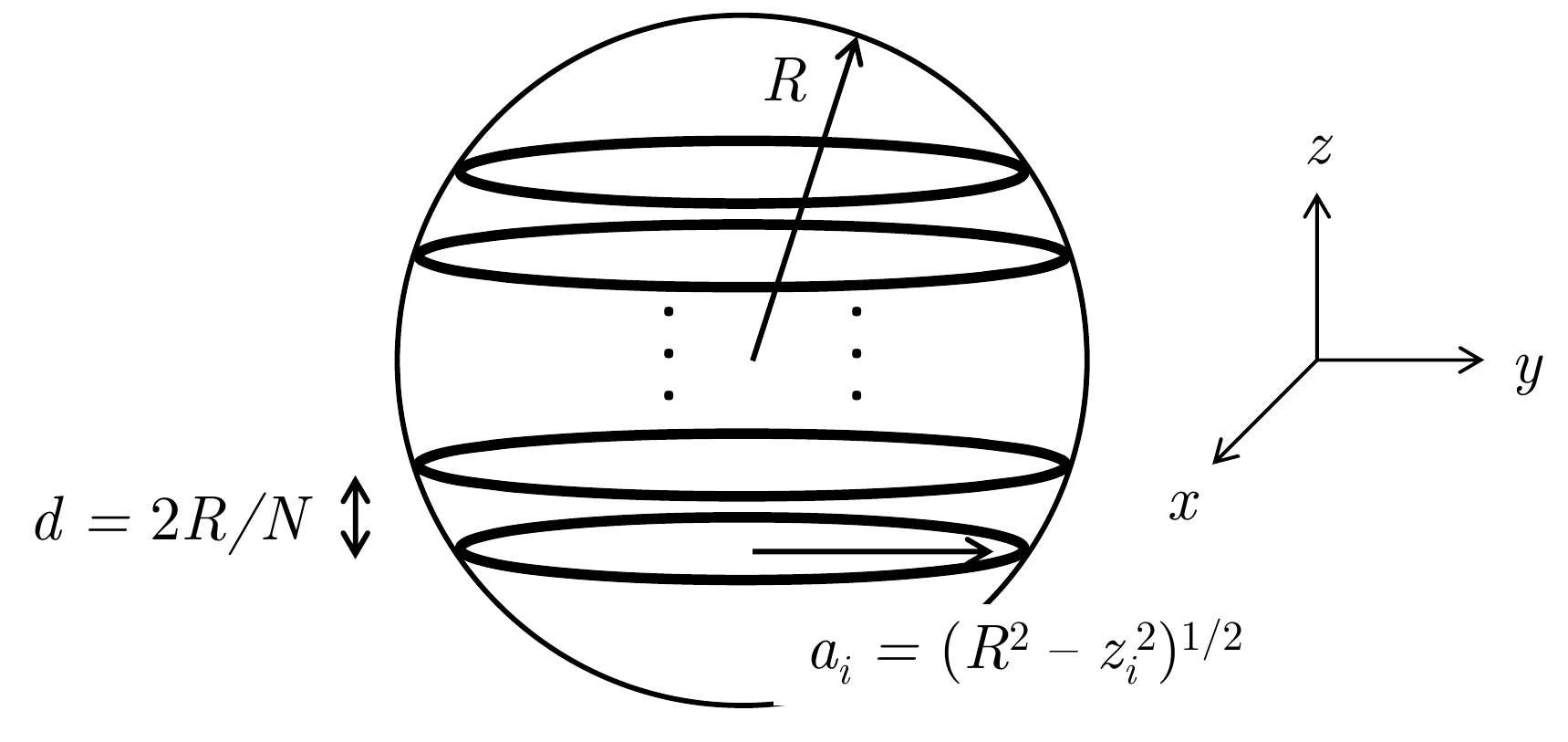}
\caption{Schematic diagram of our model for a standard discretized
spherical coil.  The thick lines indicate the circular current loops.}
\label{fig:spherical_coil_discretized}
\end{center}
\end{figure}

Finally, a schematic diagram of our discretized spherical coil is
shown in Fig.\ \ref{fig:spherical_coil_discretized}.  As indicated
there, our model consists of a spherical form of radius $R$ upon which
are wound $N$ circular current loops oriented in the $xy$-plane.  The
discretization of the spherical coil's $\vec{K} = K
\sin\theta\hat{\phi}$ continuous surface current distribution into $N$
discrete line wires proceeds in an identical manner to how the
discretization of a $\cos\theta$ coil current is performed in
Ref.\ \cite{bidinosti05}, and is as follows.  Suppose we desire to
divide the integral of the surface current, $\propto
\int\limits_0^{\pi} \sin\theta~R~d\theta \propto 2R$, into $N$ equal
parts, with each part bounded by the angles $[\alpha_{i-1},\alpha_i]$
for $i=1, 2, \ldots, N$, where $\alpha_0 = 0$ and $\alpha_N = \pi$.
We then approximate each of these parts carrying equal currents
$\propto 2R/N$ with a wire.  Then, let $\theta_i$ denote the angular
position of the $i$th wire,
where $\theta_i$ locates the midpoint
of the integrated surface current over the angular interval
$[\alpha_{i-1},\alpha_i]$.  Thus, per these definitions, $\theta_i$
must satisfy the condition
\begin{equation}
\int_{\alpha_{i-1}}^{\theta_i} \sin\theta~R~d\theta = 
\int_{\theta_i}^{\alpha_i} \sin\theta~R~d\theta = \frac{1}{2}\frac{2R}{N}.
\end{equation}
From this, it follows that $\cos\alpha_{i-1} - \cos\alpha_i = 2/N$,
and so starting from $\alpha_0 = 0$, we find that
$\cos\alpha_i = 1 - 2i/N$.  From the above integrals, it also follows
that $\cos\theta_i = \cos\alpha_{i-1} - 1/N$, from which it then
follows that the $z$-coordinate for the $i$th discretized wire is
\begin{equation}
z_i = R\cos\theta_i = R\left(1 - \frac{2i-1}{N}\right).
\end{equation}
This shows that a discretized approximation to a spherical coil
consists of circular current loops spaced at equal intervals of $d =
2R/N$ along the $z$-axis at positions $z_i = \pm \frac{1}{2}d$, $\pm
\frac{3}{2}d$, \ldots, $\pm \frac{N-1}{2}d$.  From this, it then
follows that the radius $a_i$ of the $i$th current loop is constrained
to be $a_i = \sqrt{R^2 - z_i^2}$, and the radii of the two smallest
loops, at positions $z_1 = -z_N = \frac{N-1}{2}d$, are $a_1 = a_N =
\sqrt{R^2 - z_{1,N}^2}$.  Again, we note that the above is for $N$
even; however, as before, $N$ odd, with a wire at $\theta=\pi/2$,
would not modify the symmetry of the problem.

To calculate the magnetic field of our discretized spherical coil
everywhere inside of the coil, we now employ $(r,\theta,\phi)$
spherical coordinates, where the origin is located at the center of
our spherical coil.  As shown in Refs.\ \cite{smythe,ferraro}, the
vector potential $A_{\phi,i}(\vec{x})$ of the $i$th circular current
loop (located at $\theta = \theta_i$) at some observation point
$\vec{x} = (r<R, \theta, \phi)$, as derived via the stream function
approach, is of the form
\begin{equation}
A_{\phi,i}(\vec{x}) = \frac{\mu_0 I}{2} \sum_{n=1}^\infty
\frac{\sin\theta_i}{n(n+1)}\left(\frac{r}{R} \right)^n P^1_n(\cos\theta_i)
P^1_n(\cos\theta).
\end{equation}
From this, it then follows that the $r_i$- and
$\theta_i$-components of the magnetic field at $\vec{x}$, as
calculated from $\vec{B} = \vec{\nabla} \times \vec{A}$ in
spherical coordinates, are then for $r < R$ of the explicit form
\begin{equation}
\left.\begin{array}{c}
B_{r,i}(\vec{x}) \\
B_{\theta,i}(\vec{x})
\end{array}\right\} =
\frac{\mu_0 I \sin\theta_i}{2R} \sum_{n=1}^\infty
\left(\frac{r}{R} \right)^{n-1} P^1_n(\cos\theta_i)
\left\{\begin{array}{l}
P_{n}(\cos\theta) \\
\displaystyle{-\frac{1}{n}}P^1_n(\cos\theta),
\end{array}\right.
\end{equation}
where $P_n$ denotes the ordinary Legendre polynomial and $P_n^1$
the associated Legendre polynomial.
We then summed the field components from all of the current loops
comprising the spherical coil, and then converted the
resulting $B_r(\vec{x})$ and $B_\theta(\vec{x})$ field
components in spherical coordinates to rectangular field components,
ultimately for comparison with the other two coil types.

\section{Comparisons of Field Uniformities and Field Gradients}
\label{sec:results}

\begin{figure}
\begin{center}
\includegraphics[scale=0.62]{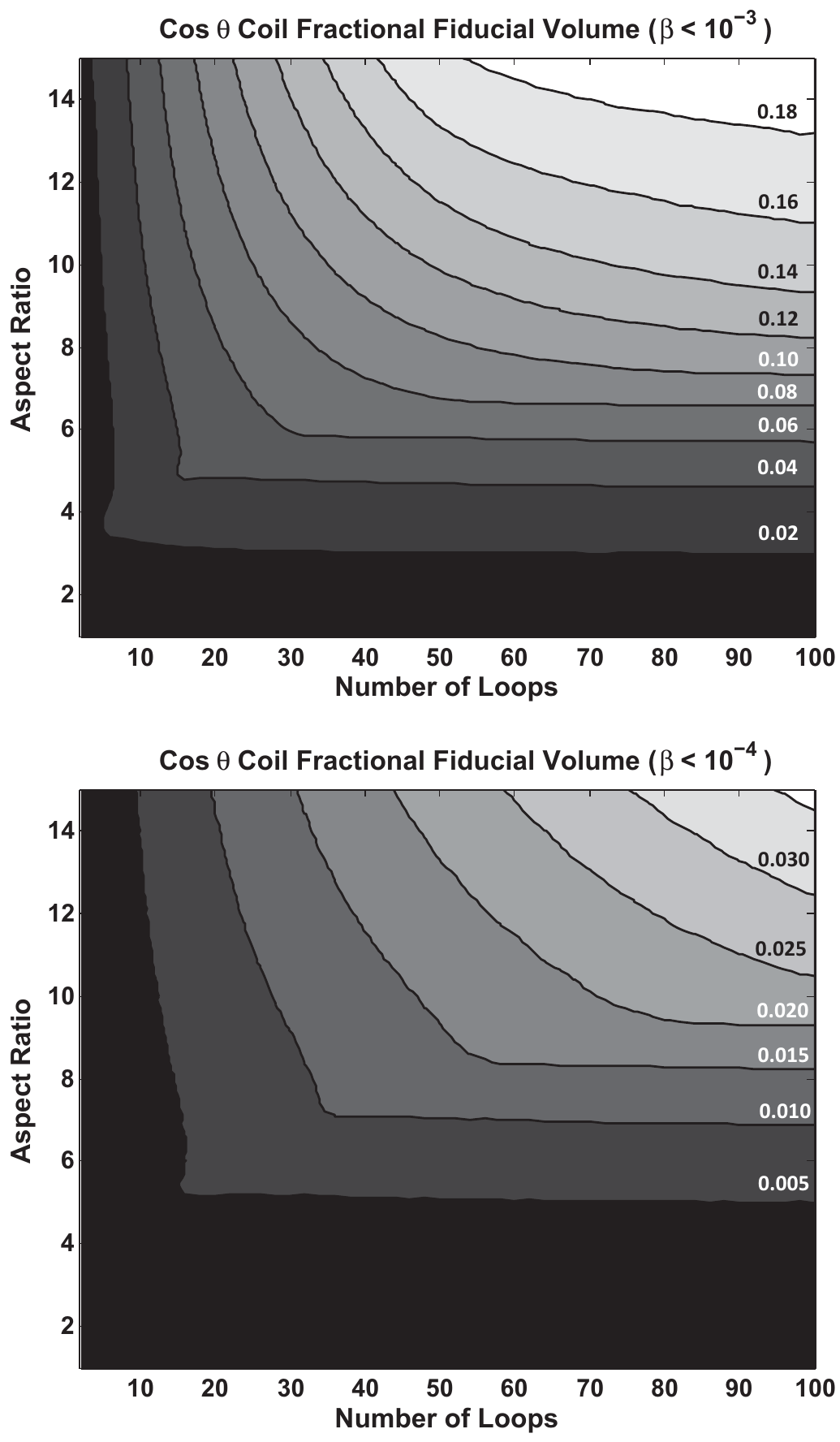}
\caption{Fractional fiducial volume for $\beta < 10^{-3}$ (top plot)
  and $<10^{-4}$ (bottom plot) field uniformities for a standard
  discretized and finite-sized $\cos\theta$ coil, as a function of the
  number of rectangular loops and the coil's aspect ratio, or
  length-to-radius ratio $L/a$.}
\label{fig:cos_theta_fractional_fiducial}
\end{center}
\end{figure}

\begin{figure}
\begin{center}
\includegraphics[scale=0.62]{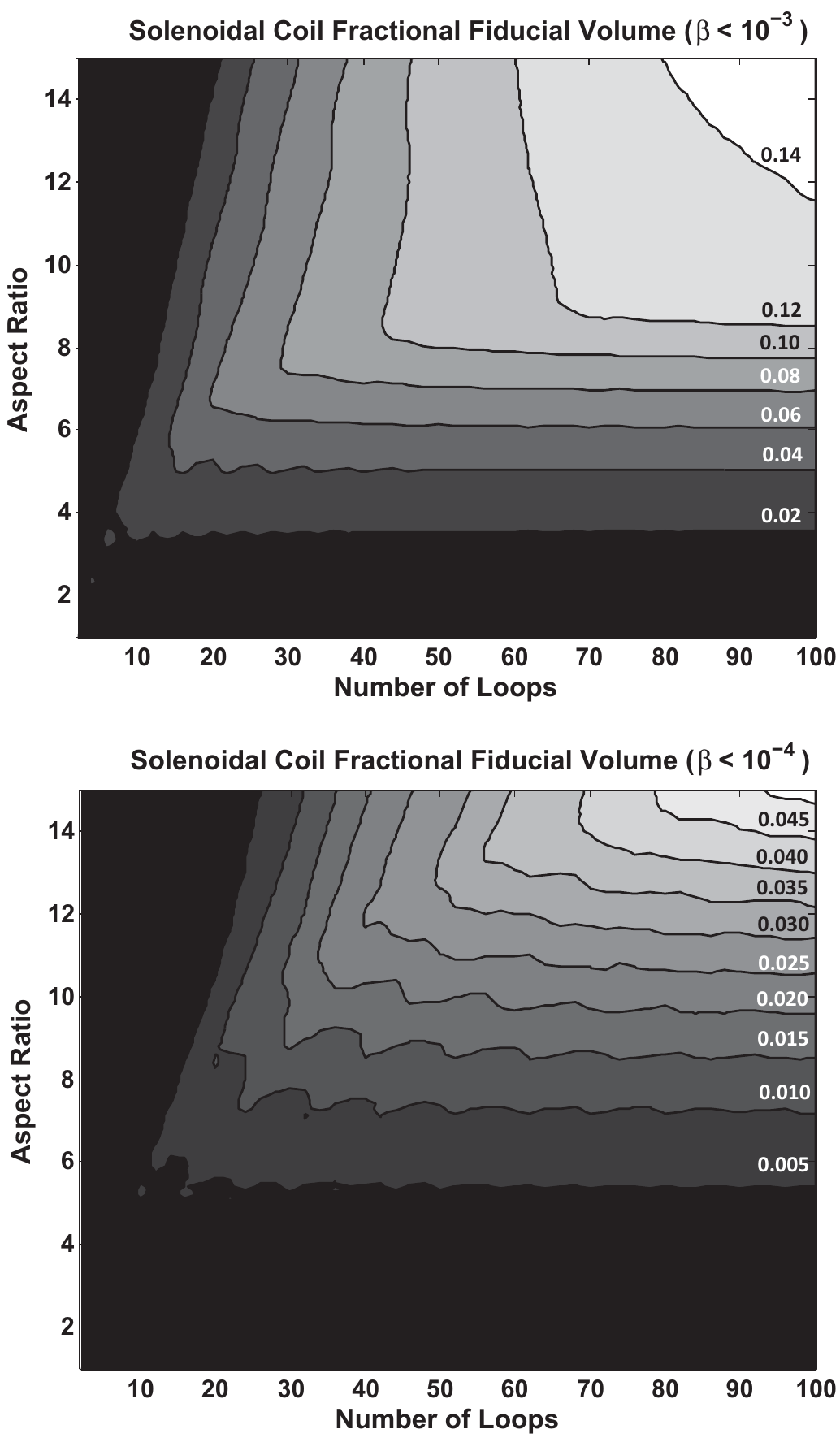}
\caption{Fractional fiducial volume for $\beta < 10^{-3}$ (top plot)
  and $\beta < 10^{-4}$ (bottom plot) field uniformities for a
  standard discretized and finite-sized solenoidal coil, as a function
  of the number of circular loops comprising the solenoid and the
  coil's aspect ratio, or length-to-radius ratio $L/a$.}
\label{fig:solenoidal_fractional_fiducial}
\end{center}
\end{figure}

\begin{figure}
\begin{center}
\includegraphics[scale=0.50]{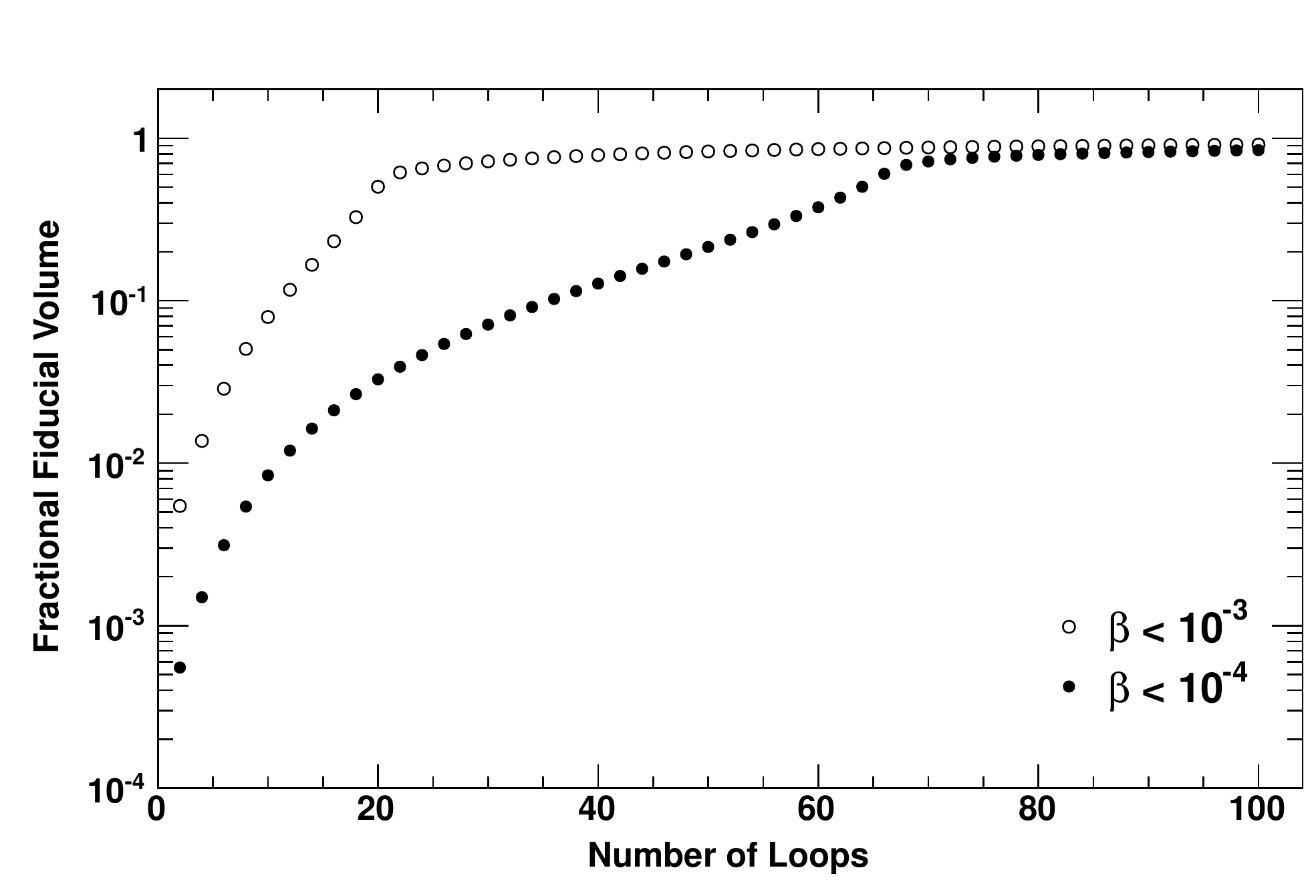}
\caption{Fractional fiducial volume for $\beta < 10^{-3}$ (open
  circles) and $\beta < 10^{-4}$ (filled circles) field uniformities
  for a standard discretized spherical coil, as a function of the
  number of circular loops comprising the spherical coil.}
\label{fig:fractional_fiducial_spherical}
\end{center}
\end{figure}

We now compare the field uniformities and gradients from standard
discretized and finite-sized $\cos\theta$, solenoidal, and spherical
coils for two example scenarios.  First, to illustrate the fiducial
volume properties of these coils, we show in
Figs.\ \ref{fig:cos_theta_fractional_fiducial}--\ref{fig:fractional_fiducial_spherical}
calculations of each coil's fractional fiducial volume (i.e., the
fraction of each coil's volume which satisfies the fiducial volume
requirement) for $10^{-3}$ and $10^{-4}$ field uniformities,
where we define the field uniformity $\beta$ at some point $\vec{x}$
relative to the center of the coil, $\vec{x}=0$, to be
\begin{equation}
\beta \equiv \left| \frac{|\vec{B}(\vec{x})| - |\vec{B}(\vec{x}=0)|}
{|\vec{B}(\vec{x}=0)|} \right|.
\end{equation}
The calculations for the $\cos\theta$ coil are shown in
Fig.\ \ref{fig:cos_theta_fractional_fiducial} for $\beta < 10^{-3}$
and $< 10^{-4}$ in two-dimensional parameter space of the number of
rectangular current loops, $N$, and the coil's length-to-radius ratio,
$L/a$.  Those for the solenoidal coil are shown in
Fig.\ \ref{fig:solenoidal_fractional_fiducial} for $\beta < 10^{-3}$
and $< 10^{-4}$ in two-dimensional parameter space in terms of the
number of circular loops, and the solenoid's length-to-radius ratio,
$L/a$.  Finally, those for the spherical coil are shown in
Fig.\ \ref{fig:fractional_fiducial_spherical} for $\beta <
10^{-3}$ and $< 10^{-4}$ as a function of the number of turns, which
is the only free parameter for a standard spherical coil of fixed
radius.

\begin{figure}
\begin{center}
\includegraphics[scale=0.68]{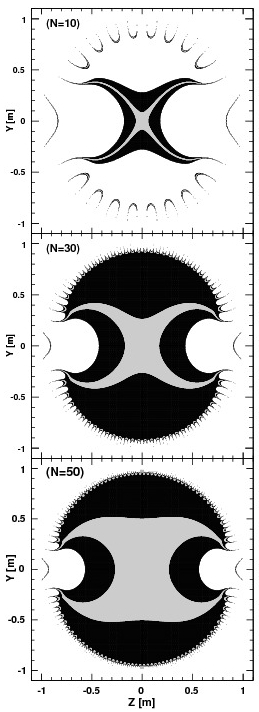}
\caption{Visualization of the fractional fiducial volumes of 1.0-m
  radius spherical coils consisting of $N=10$, 30, and 50 circular
  loops which satisfy $\beta < 10^{-3}$ (dark regions) and
  $<10^{-4}$ (gray regions) field uniformities.  Note that the circular
  current loops are in the $xy$ plane.}
\label{fig:fractional_fiducial_spherical_growth}
\end{center}
\end{figure}

We note that for the $\cos\theta$ and solenoidal coils, further
increases in the number of turns $N$ beyond the upper range of $N=100$
shown in
Figs.\ \ref{fig:cos_theta_fractional_fiducial}--\ref{fig:solenoidal_fractional_fiducial}
does not lead to appreciable increases in their fractional fiducial
volumes.  Thus, our results clearly indicate that the fractional
fiducial volume of a spherical coil can be significantly larger (by up
to an order of magnitude for spherical coils consisting of $> 60$
circular loops) than those of $\cos\theta$ and solenoidal coils.  As a
visual illustration of the uniformity properties of a spherical coil,
Fig.\ \ref{fig:fractional_fiducial_spherical_growth} shows the
fiducial volumes of 1.0-m radius spherical coils consisting of $N=10$,
30, and 50 circular loops which satisfy a $\beta <10^{-3}$ and $< 10^{-4}$
field uniformity requirement.

\begin{table}[t!]
\begin{center}
\caption{Geometry parameters for spherical coils (number of turns $N$,
radius of sphere $R$, and the resulting radii of the two smallest
current loops $a_{\mathrm{min}} \equiv a_1 = a_N$; see text in Section
\ref{sec:models_spherical}) which would satisfy a requirement on the
fractional field gradient of $\gamma < 10^{-5}$ everywhere inside of a
$10 \times 10 \times 10$ cm$^3$ cube.  These are compared with
geometry parameters (number of turns $N$, radius of cylindrical coil
form $a$, and length of cylindrical coil form $L$) for benchmark
$\cos\theta$ and solenoidal coils.  The total volume $V$ occupied by
each coil form is also listed.}
\begin{tabular}{lcccc} \hline\hline
Coil& $N$& $R$ [cm]& $a_{\mathrm{min}}$ [cm]& $V$ [m$^3$] \\ \hline
Spherical& 20&  87& 27.2& 2.76 \\
Spherical& 40&  45& 10.0& 0.38 \\
Spherical& 60&  31& 5.64& 0.12 \\
Spherical& 80&  24& 3.78& 0.06 \\
Spherical& 100& 20& 2.82& 0.03 \\ \hline
Coil& $N$& $a$ [cm]& $L$ [cm]& $V$ [m$^3$] \\ \hline
$\cos\theta$& 100& 19& 229& 0.26 \\
Solenoidal& 100& 13& 249& 0.13 \\ \hline
\end{tabular}
\label{tab:examples_geometry_parameters_cube}
\end{center}
\end{table}

As a second example, we consider an experimental requirement on the
fractional field gradient, $\gamma$, which we define to be
\begin{equation}
\gamma \equiv \frac{1}{|\vec{B}(\vec{x}=0)|}
\left|\frac{\partial B_i}{\partial x_i}\right|,
\end{equation}
where $B_i$ denotes the field component along the field's primary
direction (i.e., the $x$-component for the $\cos\theta$ coil and the
$z$-component for the solenoidal and spherical coils).  To illustrate,
suppose a hypothetical requirement is that $\gamma < 10^{-5}$
everywhere inside of a $10 \times 10 \times 10$ cm$^3$ rectangular
volume located at the center of each coil.  In Table
\ref{tab:examples_geometry_parameters_cube} we show examples of
spherical coil geometry parameters which would satisfy this
requirement, and compare these with benchmark examples of geometry
parameters for approximate best-case $\cos\theta$ and solenoidal coils
(i.e., large length-to-radius ratios and number of turns, as suggested
by the fractional fiducial volume calculations presented in
Figs.\ \ref{fig:cos_theta_fractional_fiducial}--\ref{fig:solenoidal_fractional_fiducial}).
Our results indicate that a particular requirement on the fractional
field gradient could be achieved with a spherical coil occupying a
smaller volume than the volumes occupied by $\cos\theta$ and
solenoidal coils.

However, we do note that a potential appeal of the $\cos\theta$-coil
(if wound with a ``saddle winding'' \cite{bidinosti05,perezgalvan11})
and solenoidal-coil geometries is that their winding patterns permit
significant access to their interior regions along their respective
$z$-axes (i.e., via a circular aperture of radius equal to the radius
$a$ of their cylindrical form), whereas access to the interior of a
spherical coil along its $z$-axis is constrained by the radii of the
two smallest current loops, $a_1 = a_N = \sqrt{R^2 - z_{1,N}^2}$,
located at positions $z_1 = -z_N = \frac{N-1}{2}d$, as discussed
previously in Section \ref{sec:models_spherical}.  If a $\cos\theta$
coil is wound with rectangular loops (as in our model calculations),
access to the interior along its $z$-axis is restricted by the
spacing $d=2a/N$ between adjacent rectangular loops (see
Fig.\ \ref{fig:cos_theta_coil_discretized}).

For access to their interior regions along the transverse direction,
the dimensions of the $\cos\theta$ coil's two smallest rectangular loops
are $\frac{2 a}{N} \sqrt{2 N - 1}$ ($y$-direction) $\times$ $L$
($z$-direction; see Fig.\ \ref{fig:cos_theta_coil_discretized}).  The
solenoidal coil can, in principle, be accessed in the transverse
direction via the gaps of size $\frac{L}{N-1}$ in between adjacent
circular windings, and the spherical coil can also be accessed in the
transverse direction in between the circular current loops, which are
spaced at equal intervals of $d=2R/N$ along its $z$-axis (see
Fig.\ \ref{fig:spherical_coil_discretized}).

\section{Summary}
\label{sec:summary}

In this paper we have presented a model for numerical calculations of
the magnetic field of a standard discretized spherical coil.  Because
a discretized spherical coil does not suffer from any ``finite-length
effects'' that are inherent to standard discretized and finite-sized
$\cos\theta$ and solenoidal coils, our calculations indicate that the
fractional fiducial volume of a discretized spherical coil is
potentially significantly larger than those of standard (i.e.,
non-optimized) discretized and finite-length $\cos\theta$ and
solenoidal coils. \\

\noindent\textbf{Acknowledgments} \\ \\
\indent This work was supported in part by the U.\ S.\ Department of
Energy Office of Nuclear Physics under Award Number DE-FG02-08ER41557.
We thank Prof.\ J.\ W.\ Martin for encouraging us to write this paper,
and Prof. C.\ B.\ Crawford for several valuable discussions.  We also
thank an anonymous referee for many excellent suggestions which, we
believe, significantly improved this paper.


\end{document}